\documentclass[aps,prl,secnumarabic,twocolumn,groupedaddress,showpacs,amsmath,amssymb]{revtex4}

\usepackage{epsfig}
\usepackage{amssymb}
\usepackage{amsmath}
\usepackage{graphicx}
\usepackage{amsfonts}
\usepackage{epsfig}
\usepackage{bm}
\usepackage{amsmath}
\usepackage{amssymb}
\usepackage{latexsym}

\begin{document}

\title{Hydrodynamic instability during non-uniform growth of a helium crystal}

\author{S. N. Burmistrov}
\email[]{burmi@kurm.polyn.kiae.su} \affiliation{Institute of Superconductivity and Solid State Physics,
Kurchatov Institute, 123182 Moscow, Russia}
\author{L. B. Dubovskii}
\affiliation{Institute of Superconductivity and Solid State Physics, Kurchatov Institute, 123182 Moscow,
Russia}
\author{V. L. Tsymbalenko}
\affiliation{Institute of Superconductivity and Solid State Physics, Kurchatov Institute, 123182 Moscow,
Russia}


\begin{abstract}
We analyze an analog of the  hydrodynamic Rayleigh-Taylor instability for the liquid-solid phase interface
under non-uniform growth of the solid phase. The development of the instability starts on conditions of an
accelerated interface growth and if the magnitude of acceleration exceeds some critical value. The plane and
spherical shapes of the interface are considered. The observation of the instability can be expected for
helium crystals in the course of their abnormal fast growth.
\end{abstract}


\maketitle

{\bf Introduction.} The fast growth kinetics of the crystal-superfluid helium interface results in appearing
the instabilities similar to those inherent in the interfaces between fluid media. For instance, it is shown
that the crystal surface is unstable in the steady tangential flow of a fluid \cite{Ka,No}, demonstrating thus
the Kelvin-Helmholtz type of instability. The phenomenon is observed qualitatively as a distortion of the
crystal surface in the fluid jet \cite{Ma}. The onset of such instability occurs when the harmonics increasing
exponentially in time appear in the spectrum of crystallization waves.
\par
\begin{figure}
\includegraphics[scale=2.0]{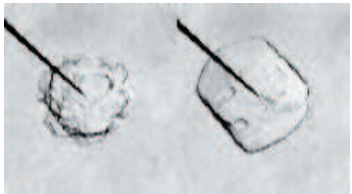}
\caption{The growth of a crystal at 0.47 K and initial overpressure 5.2 mbar. The left frame corresponds to
0.19~ms after the crystal nucleation. The right frame is taken at 80~ms after the crystal nucleation at the
pressure close to the phase equilibrium pressure. The vertical size of the frames is 2.4~mm.} \label{fig1}
\end{figure}

It is known after \cite{No} that the wave spectrum does not vary in the case of the normal incidence of the
fluid onto the surface. Hence the growth of a crystal should not lead to an instability of its surface.
However, in the experiment on the free growth of the crystal initiated at the needle point immersed into the
liquid bulk \cite{Ts,Tsy} one has observed an instability at the initial stage of the abnormal fast growth.
After nucleation the crystal grows first with the clear kinetic prism-shaped faceting. Then, if the initial
overpressure at which the crystal is nucleated is sufficiently large, the crystal after 0.1--0.2~ms takes the
shape close to the spherical one with a sizeable ripple on the surface, see Fig.~\ref{fig1}. The magnitude of
the threshold overpressure lies within 6--8 mbar and the threshold of the instability does not depend upon the
temperature within errors \cite{Tsym}.
\par
The attempts to explain this phenomenon within the known theoretical works have been unfruitful \cite{Tsym}.
The essential point, which differs the experiment from the cases considered theoretically, is unsteady growth
conditions. Under such large overpressures the crystal grows in the oscillating regime with the typical
oscillation period $\sim$0.4~ms and reaches the size $\sim$1 mm for the first half-cycle. The linear velocity
for the motion of the crystal boundary runs up to $\sim$10 m/s. The interface-directed velocity of the liquid
phase is one order of the magnitude as smaller due to difference in the densities of the liquid and solid
phases. During the first half-cycle the crystal growth rate varies from maximum to zero one. This gives an
estimate $\sim$10$^5$ m/s$^2$ for the acceleration of the interface and $\sim$10$^4$ m/s$^2$ for the fluid
beside the crystal. Another essential point here is a direction of acceleration which is opposite to the
velocity of the crystal growth. In the non-inertial reference frame related to the interface segment this
results in the acceleration directed from the crystal to the liquid. Since the crystal density is larger than
the liquid one, the situation proves to be looking like the normal Rayleigh-Taylor instability \cite{T,B} for
the interface between two classical immiscible fluids of various densities. The interface is unstable or
stable according to whether  the acceleration is directed from the lighter to the heavier fluid or vice versa.
\par
However, the fast growth kinetics of a helium crystal in the abnormal state gives rise to a series of
qualitative features. The classical Rayleigh-Taylor instability develops at the fixed volume of each phase. In
contrast, the impetuous growth of the liquid-crystal interface can result in varying the interface profile due
to phase conversion. In addition, besides the kinetic liquid and interface surface energies it is necessary to
involve the energy of phase conversion depending on the imbalance from the phase equilibrium. Thus, in spite
of  superficial resemblance such instability has specific features which differ it from the familiar
Rayleigh-Taylor instability.
\par
The variation of the surface shape of a crystal at its non-uniform growth has been studied only in the
experiments on abnormal crystal growth \cite{Tsy}. The classical instability of the crystal-liquid helium
interface in the field of gravity is observed by Demaria, Lewellen, and Dahm \cite{De}. In these experiments a
cell in which the solid and liquid phases occupy initially the lower and upper halves, respectively, is
inverted mechanically by 180$^{\circ}$. After inversion the crystal starts to melt, the interface descending
along the walls and ascending in the centre of a cell. The authors gave only the schematic description of the
process without concrete spatial scale and time reference points over technical failures with determining the
interface position from the filming of the process. Unfortunately, the temperatures at which the experiments
are performed are not indicated with the exception of the crystal growth temperature 1.1~K and maximum one
1.7~K. The question about the effect of roughening transitions upon stability of the crystal facet in the
field of gravity has remained open. We note only the following. The stability observed for the shape of the
crystal grown at the needle point \cite{Ts} with its lower facet under condition favorable for developing the
Rayleigh-Taylor instability evidences that the acceleration of gravity did not result in any instability of
the atomically smooth interface for the sizes not exceeding 10~mm at least.
\par
In the present work we give a theory of the phase interface instability for the case of the non-uniform
interface growth rate. A comparison is presented between the calculation results and the experimental data on
developing the instability of the crystal shape at the abnormal growth.
\par
{\bf The growth kinetics of the crystal surface.} Let the plane liquid-solid interface grow at the rate
$V=\dot{L}(t)$ and regions $z>L(t)$ and $z<L(t)$ be occupied with the superfluid and the solid $^4$He,
respectively. We consider stability of the growing interface with respect to its small distortions from the
plane shape, i.e., interface $z$-coordinate is taken as $Z=L(t)+\zeta(x, y,\, t)$. Assuming the crystal growth
rates to be small compared with the first and second sound velocities, we consider the hydrodynamics of a
superfluid in the approximation of incompressible liquid and constancy of the entropy density. In this case
\cite{LL} the equations for the normal and superfluid motions can be separated and the potentials of the
normal and superfluid velocities satisfy, respectively, $\nabla^2 \phi _n=0$ and $\nabla^2 \phi _s=0$. Thus,
the solutions with wave vector $\bm{q}=(q_x,\, q_y)$ parallel to the surface can be represented as
\begin{gather}\nonumber
\phi _s=u_s(t)z+A_s(t)\exp (i\bm{q}\bm{r}-qz)
\\
\phi _n=u_n(t)z+A_n(t)\exp (i\bm{q}\bm{r}-qz)\, ,\nonumber
\end{gather}
where $\bm{r}=(x,y)$ and velocities $u_s$  and $u_n$ correspond to those of the undisturbed motion. The
pressure in the fluid is a sum of pressures $P=P_n+P_s$ and
\begin{gather}\nonumber
P_s=P_{s\,\infty}-\rho _s\bigl(\dot{\phi}_s+(\nabla\phi _s)^2/2\bigr)+\rho _sg(L_{\infty}-z)
\\
P_n=P_{n\,\infty}-\rho _n\bigl(\dot{\phi}_n+(\nabla\phi _n)^2/2\bigr)+\rho _ng(L_{\infty}-z)\, .\nonumber
\end{gather}
\par
Let us employ the boundary conditions used in the theory of crystallization waves and crystal growth kinetics.
From the continuity of the mass flow we have at the interface $z=Z(x, y,\, t)$
\begin{equation}\nonumber
 j_{\nu}=\rho
_nV_{n\,\nu} +\rho _sV_{s\,\nu} =(\rho -\rho ')\dot{Z}\, .
\end{equation}
Here $\bm{\nu}$ is the normal to the surface.
\par
To simplify the consideration, farther on we neglect shear components $\sigma _{i\neq k}$ for the stress
tensor $\sigma _{ik}$ in the solid phase. In other words, we suppose the stress tensor to be diagonal, i.e.,
$\sigma _{ik}=-P'\delta _{ik}$. Then the solid looks like a liquid under pressure $P'$.
\par
From the continuity of the momentum flux density and involving the surface tension and interfacial curvature,
we arrive at
\begin{gather}\nonumber
P+\rho _n(\bm{v}_n-\dot{Z}\bm{\nu})^2+\rho _s(\bm{v}_s-\dot{Z}\bm{\nu})^2-(P'+\rho '\dot{Z}^2)=
\\
 =\gamma _{ik}\partial Z ^2/\partial x_i\partial x_k\, ,\;\;\;\;\; \nonumber
\end{gather}
where $\gamma _{ik}=\alpha\delta _{ik} +\partial\alpha ^2/\partial\varphi _i\partial\varphi _k$ is the surface
stiffness tensor expressed in terms of surface tension $\alpha$ and derivatives over angle $\varphi$ between
the normal to the displaced surface  and the axis $z$. The quantities with dash are referred to the solid
phase.
\par
For the normal velocity according to work \cite{AK}, we suppose the sticking of the normal component to the
surface  as for a viscid fluid
\begin{equation}\nonumber
v_{n\,\nu}=\dot{Z}\, .
\end{equation}
\par
The growth rate of the solid phase is described by the kinetic growth coefficient $K$ and connected with the
difference in chemical potentials  per unit mass as usual
\begin{gather}
\dot{Z}=K\left[\mu +\frac{(\bm{v}_s-\dot{Z}\bm{\nu})^2}{2}-\left(\mu '+\frac{\dot{Z}^2}{2}\right)\right]\, .
\nonumber
\end{gather}
Then we have for the difference in the chemical potentials and temperature \cite{LL}
\begin{gather}
\;\;\;\;\;\;\; \mu -\mu ' = \nonumber
 \\ =  \sigma (T-T_{\infty})+\frac{P-P_c}{\rho}-\frac{\rho
_n}{\rho}\,\frac{(\bm{v}_n-\bm{v}_s)^2}{2}-\,\frac{P'-P_c}{\rho '}\, , \nonumber
\\
 T-T_{\infty}
 =\! \frac{\rho _n}{\sigma\rho}\left(\frac{P_n-P_{n\,\infty}}{\rho _n}-\frac{P_s-P_{s\,\infty}}{\rho
_s}-\frac{(\bm{v}_n-\bm{v}_s)^2}{2}\right)\nonumber
\end{gather}
where $\sigma$ is the entropy and the quantities with index \textquotedblleft
 $\infty$\textquotedblright\, stand for the magnitudes taken far from the interface. Pressure $P_c$
denotes the equilibrium pressure at which the phase transition takes place.
\par
Solving a set of the boundary conditions, one finds that the undisturbed motion of the plane interface at rate
$V=\dot{L}(t)$ obeys
\begin{equation}
V\frac{\rho '}{K}=\frac{\rho '-\rho}{\rho}\, \bigl[\Delta P+\rho g(L_{\infty}-L)\bigr] + \rho
_{\text{ef}}\!\left(\dot{V}L+\frac{V^2}{2}\right) .\nonumber
\end{equation}
Here $\rho _{\text{ef}}$ is the effective density of the interface
\begin{eqnarray*}
\rho _{\text{ef}}=\rho _n +(\rho '-\rho _s)^2/\rho _s
\end{eqnarray*}
and $\Delta P=P_{\infty}-P_c$ is the overpressure. The equation for the disturbed motion in the linear
approximation reads
\begin{eqnarray*}
\rho _{\text{ef}}\,\frac{\ddot{\zeta}}{q}+\frac{\rho '}{K}\,\dot{\zeta}+\bigl[\gamma _{ik}q_iq_k +(\rho
'-\rho)g-\rho _{\text{ef}}\dot{V}\bigr]\,\zeta =0 \, .
\end{eqnarray*}
For the uniform growth of the interface $\dot{V}=0$, the equation goes over into the familiar dispersion
relation for the crystallization wave spectrum \cite{AK,BAP}.
\par
{\bf The instability.} Let us consider the stability of the flat-shaped interface with respect to small
perturbations $\zeta \sim\exp (\lambda t)$ for the uniformly accelerated growth of the crystal. The root with
$\text{Re}\,\lambda (q)>0$ corresponds to instability. For the acceleration exceeding the threshold one
$\dot{V}_c=g(\rho '-\rho)/\rho _{\text{ef}}$, the interfacial distortion will increase for the wave vectors
satisfying $\gamma _{ik}q_iq_k<\rho _{\text{ef}}\dot{V}-(\rho '-\rho)g$. Note that $\dot{V}_c$ does not depend
on the growth coefficient, i.e., on the dissipative properties of the interface, and is positive. The latter
corresponds to the case when the interfacial acceleration is directed to the fluid. Thus, the stability due to
non-uniform growth of the plane interface appears only for the accelerated growth of a crystal.
\par
The value $q_0$ corresponding to the maximum magnitude  $\text{Re}\,\lambda (q)$ meets the shortest time for
the development of the instability which will be characterized by the spatial scale $2\pi /q_0$. The value
$q_0$ can be found from the equation
\begin{eqnarray*}
q_0^2=\frac{q_c^2}{3}-\,\frac{\sqrt{2}}{3}\,\frac{\rho '}{\sqrt{\gamma\rho _{\text{ef}}}}\,\frac{q_0^{3/2}}{K}
\end{eqnarray*}
where $q_c$ is the value related to the upper bounds of instability according to $\gamma q_c^2=\rho
_{\text{ef}}\dot{V}-g(\rho '-\rho)$. For the large magnitudes of the growth coefficient or large acceleration
while $K^4\dot{V}\gg\rho ^{\prime\, 4}/\gamma\rho _{\text{ef}}^3\,$, values $q_0$ and $\lambda (q_0)$ are
equal approximately to
\begin{gather*}
q_0=\frac{q_c}{\sqrt{3}}\left(1-\,\frac{\rho '}{(6\sqrt{3}\,\gamma\rho _{\text{ef}}\,K^2q_c)^{1/2}}\right) ,
\\
\lambda _0=\left(\frac{2\gamma}{3\sqrt{3}}\,\frac{q_c^3}{\rho _{\text{ef}}}\right)^{1/2}
-\,\frac{q_c}{2\sqrt{3}}\,\frac{\rho '}{\rho _{\text{ef}}\, K}\,  .
\end{gather*}
In the opposite limit $K^4\dot{V}\ll\rho ^{\prime\, 4}/\gamma\rho _{\text{ef}}^3$ one has roughly
\begin{gather*}
q_0=\left(\frac{\gamma\rho _{\text{ef}}}{\rho ^{\prime\, 2}}\, K^2q_c^4\right)^{\! 1/3}  ,
\\
\lambda _0=\frac{K}{\rho '}\,\gamma q_c^2 \, .
\end{gather*}
\par
On the whole, values $q_0$ and $\lambda _0$ decrease as the kinetic growth coefficient reduces or dissipation
of the interface enhances. From the experimental point of view this may require the crystal surface of
sufficiently large sizes $d>2\pi /q_0$ and the large time of supporting the accelerated growth $t>1/\lambda
_0$ in order to realize the interfacial instability at the uniformly  accelerated   growth.
\par
{\bf The spherical geometry.} In the experiment a crystal nucleates at the needle point and then grows free.
The anisotropy of the kinetic growth coefficient is not large. A ratio of the maximum to minimum growth rate
does not exceed 2--3. Let us consider the stability of the spherical shape of the growing crystal. For
simplicity, we assume the isotropy of the surface tension $\alpha$. As it follows from the estimates given in
the Introduction, the acceleration of gravity is negligibly small as compared with the typical magnitudes of
the interfacial acceleration.
\par
The equation of the interface is given as $r=R_s(t,\, \Omega )=R(t)+\zeta (t,\, \Omega )$ where $\zeta =\zeta
_l(t)Y_l(\Omega )$ is a fluctuation of the interface expanded in the spherical harmonics of degree $l=0,\,1,\,
2\ldots$. We seek for the velocity potentials of the normal $\bm{v}_n=\nabla\phi _n$ and superfluid
$\bm{v}_s=\nabla\phi _s$ motions in the form
\begin{gather*}
\phi _s =-u_sR^2/r\, +A_lY_l/r^{l+1}
\\
\phi _n =-u_nR^2/r\, +B_lY_l/r^{l+1}\, .
\end{gather*}
Using the same boundary conditions as above, we find the velocities of the undisturbed flow of the liquid
phase
\begin{eqnarray*}
 u_s=\dot{R}\, (\rho _s-\rho ')/\rho _s\, , \;\;\;\;\;\;\;\; u_n=\dot{R}\, ,
\end{eqnarray*}
and coefficients $A_l$ and $B_l$ describing the disturbed motion of the interface
\begin{gather*}
A_l=-\,\frac{\rho _s-\rho '}{\rho _s}\,\frac{R^{l+2}}{l+1}\left(\dot{\zeta}+\frac{2\dot{R}}{R}\,\zeta\right)
\\
B_l=-\,\frac{R^{l+2}}{l+1}\left(\dot{\zeta}+\frac{2\dot{R}}{R}\,\zeta\right)\, .
\end{gather*}
Employing relation for the pressure and the dependence between growth rate and chemical potential difference,
we obtain for the undisturbed growth of the solid phase
\begin{equation}\nonumber
\rho _{\text{ef}}\!\left(\! R\ddot{R}+\frac{3}{2}\,\dot{R}^2\!\right)\! +\frac{\rho '}{K}\,\dot{R}=\frac{\rho
'-\rho}{\rho}\!\left(\!\Delta P-\frac{\rho}{\rho '-\rho}\,\frac{2\alpha}{R}\right) .
\end{equation}
The growth equation looks like the motion of a particle with  the effective mass  $M(R)=4\pi\rho
_{\text{ef}}R^3$, drag force $4\pi R^2\rho 'K^{-1}\dot{R}$, and potential energy $U(R)=4\pi\alpha R^2-((\rho
'-\rho)/\rho )\Delta P(4\pi R^3/3)$.
\par
The small deformations for the spherical surface of the solid phase are described by the relation
\begin{gather}\nonumber
\rho _{\text{ef}}\!\left(\frac{R}{l+1}\,\ddot{\zeta}_l+\frac{3}{l+1}\,\dot{R}\dot{\zeta}_l\right) +\frac{\rho
'}{K}\,\dot{\zeta}_l +
\\
+\left(\alpha\,\frac{(l-1)(l+2)}{R^2}-\rho _{\text{ef}}\,\frac{l-1}{l+1}\,\ddot{R}\right)\zeta _l =0  .
\label{f11a}
\end{gather}
For $\dot{R}=\ddot{R}=0$ and in the lack of energy dissipation, the equation describes the spectrum of
crystallization waves on the spherical interface
 $$ \omega _{0\, l}^2=\frac{\alpha}{\rho _{\text{ef}}R^3}\, (l-1)(l+1)(l+2) \, .
 $$
\par
The asymptotic behavior of function $\zeta _l (t)$ at large $t$ is given by
 $$
\zeta _l (t)\sim \exp\!\!\left(\int\limits _{\: 0}^{t}\! dt'\left[- \lambda _l(t')+\sqrt{\lambda _l^2
(t')+\dot{\lambda}_l(t')-\nu _l (t')}\,\right]\!\right)
 $$
where
\begin{eqnarray*}
\lambda _l(t) & = & \frac{3}{2}\,\frac{\dot{R}}{R} +\frac{1}{2}\,\frac{\rho '}{\rho
_{\text{ef}}}\,\frac{l+1}{KR}\, ,
 \\
\nu _l(t) & = & \omega _{0\, l}^2-(l-1)\,\ddot{R}/R\,  .
\end{eqnarray*}
\par
\begin{figure}
\includegraphics[scale=1]{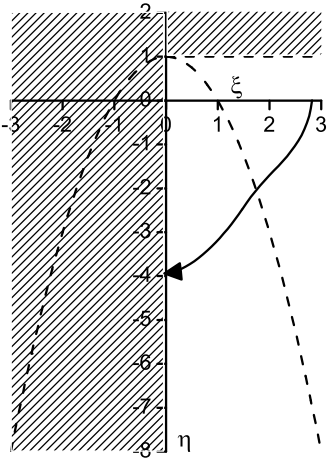}
 \caption{The phase diagram for the stability of a spherical crystal at the non-uniform growth.
The dashed parabola separates the regions of oscillating (below the curve) and relaxation modes. The regions
of instability are shown with shading. The curved arrow demonstrates schematically, but in the scale, the
trajectory of the crystal growth during the first half-cycle of the oscillating growth in the abnormal state.}
 \label{fig2}
\end{figure}

For the fixed time instant, the proper frequencies are determined by equation (\ref{f11a}). The onset of
instability at the fixed time occurs if the imaginary part of the frequency becomes positive.  To estimate, we
consider the perturbation as $\zeta _l\sim\exp (-i\omega t)$. In the dimensionless units
 $$
\Omega =\frac{\omega}{\omega _{0\, l}}\: , \; \; \xi = \frac{1}{2\omega _{0\,
l}}\left(\frac{3\dot{R}}{R}+\frac{\rho '}{\rho _{\text{ef}}}\,\frac{l+1}{KR}\right) , \;\; \eta
=\frac{l-1}{\omega _{0\, l}^2}\,\frac{\ddot{R}}{R}
 $$
the proper frequencies are determined from
 $$
\Omega _{1,\, 2}=-i\xi\pm\sqrt{1-\eta-\xi ^2}\, .
 $$
The condition of appearing the instability is $\text{Im}(\Omega)>0$. The result is given in Fig.~\ref{fig2}.
Provided that $1-\eta -\xi ^2>0$, the frequencies have the real part but the instability appears for the
negative value of parameter $\xi$. If the radicand  is negative, the both frequencies are imaginary. In this
case the left half-plane of the phase diagram corresponds also to instability. The appearance of the
instability takes place in the right half-plane for $\eta >1$.
\begin{figure}
\includegraphics[scale=0.5]{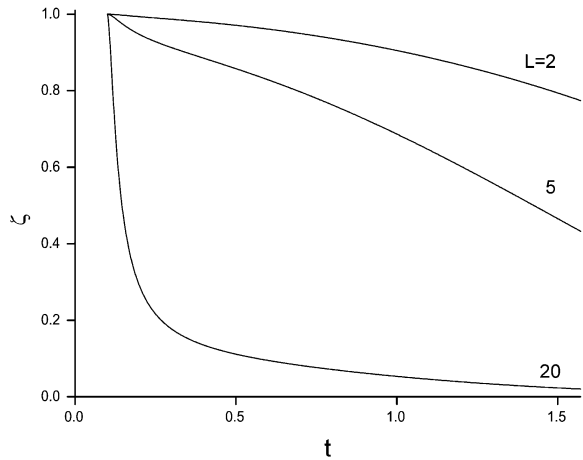}
 \caption{ The damping of the spherical harmonics during crystal growth as a function of the time
and degree of harmonic. The numbers beside the lines denote the degree of harmonic. The normalized time
$t=2\pi\! f_0\tau$ is put along the abscissa axis.}
 \label{fig3}
\end{figure}
\par
In Fig.~\ref{fig3} a simulation is shown for the behavior of spherical harmonics at the initial stage of the
abnormal crystal growth. The variation of the crystal radius is described approximately by expression $R =
R_0\sin 2\pi\! f_0\tau$. The following values $R_0 =$ 0.1~cm, $K=$~0.1 s/cm, $f_0 =$~2.5~kHz of experimental
parameters at 0.5~K and overpressure 6~mbar  are taken to calculate. The other quantities are the tabular
data. The deformation of unit amplitude is put for the initial values. The velocity is varied within (-100,
100). In the figure the solutions are given with zero initial velocity. Note that the variation of the initial
velocity in the range analyzed does not lead to the qualitative change in the behavior of the curves with the
exception of first move at which fast relaxation occurs. Next in all cases the damping of the harmonics,
checked to $l=200$, is observed. The increase of the degree of the harmonic, as it follows from the
simulation, results in the faster relaxation of the deformation. Thus, the numerical simulation agrees with
the previous conclusion about the crystal interface stability at the initial growth stage.
\par
{\bf Summary.} Non-uniformity of the crystal growth results in new qualitative hydrodynamic instability of the
liquid-solid interface. At the plane interface the development of instability starts only for the accelerated
growth of the solid phase when the magnitude of the interfacial acceleration exceeds the critical one
determined by the acceleration of gravity.
\par
For free growth of the solid phase when its shape is almost spherical, the conditions of appearing the
instability are determined both by the interfacial acceleration and by the growth rate. The experimental
situation in which one may expect an observation of such instability is realized with the abnormal helium
crystal growth when the interfacial mobility increases drastically by several orders of the magnitude
\cite{Tsymb} and the interfacial velocity and acceleration reach the huge magnitudes as is noted in the
Introduction. For lowest mode $l=2$, the normalized order-of-magnitude parameters are about $\xi\sim 10^{2}$
and $\eta\sim -10^{4}$. However, during the first half-cycle of the oscillating growth, as is seen from
Fig.~\ref{fig2}, the trajectory of the crystal growth lies in the fourth quadrant in which the instability
considered here does not appear. Thus the involvement of the non-uniform interfacial motion alone may fail to
explain the liquid-solid interface instability observed. Note that none of the hydrodynamic instabilities
known so far can be responsible for the instability of the crystal shape observed at the abnormal crystal
growth \cite{Tsym}.
\par
{\bf Acknowledgments.} The work is supported by the RFBR Grant No.~05-02-16806.


\begin{thebibliography}{99}
\bibitem{Ka} M.~Yu.~Kagan,  Zh. Eksp. Teor. Fiz. \textbf{90}, 498 (1986) [Sov. Phys. JETP \textbf{63}, 288 (1986)].
\bibitem{No} P.~Nozieres, M.~Uwaha, J. de Physique \textbf{47}, 263 (1986).
\bibitem{Ma} L.~A.~Maksimov, V.~L.~Tsymbalenko,  Zh. Eksp. Teor. Fiz.  \textbf{122}, 530 (2002)
[JETP \textbf{95}, 455 (2002)].
\bibitem{Ts} V.~L.~Tsymbalenko, Fiz. Nizk. Temp. \textbf{21}, 162 (1995) [Low Temp. Phys. \textbf{21},
120 (1995)].
\bibitem{Tsy} V.~L.~Tsymbalenko, J. Low Temp. Phys. \textbf{121}, 53 (2000).
\bibitem{Tsym} V.~L.~Tsymbalenko,  Zh. Eksp. Teor. Fiz. \textbf{126}, 1391 (2004) [JETP \textbf{99}, 1214 (2004)].
\bibitem{T} G.~I.~Taylor, Proc. Roy. Soc. (London) \textbf{A201}, 192 (1950).
\bibitem{B} G.~Birkhoff, \textit{Hydrodynamics}, (Princeton Press, 1960).
\bibitem{LL}  L.~D.~Landau, E.~M.~Lifshits, \textit{Fluid Mechanics}, (Pergamon Press, 1987), \S 140.
\bibitem{AK} A.~F.~Andreev, V.~G.~Knizhnik,  Zh. Eksp. Teor. Fiz. \textbf{83}, 416 (1982)
[Sov. Phys. JETP \textbf{56}, 226 (1982)].
\bibitem{De} C.~D.~Demaria, J.~W.~Lewellen, A.~J.~Dahm, J. Low Temp. Phys. \textbf{89}, 385 (1992).
\bibitem{BAP} S.~Balibar, H.~Alles,  A.~Ya.~Parshin, Rev. Mod. Phys. \textbf{77}, 317 (2005).
\bibitem{Tsymb} V.~L.~Tsymbalenko, J. Low Temp. Phys.  \textbf{138}, 795 (2005).

\end{thebibliography}
\end{document}